\documentclass[%
 reprint,
unsortedaddress,
 amsmath,amssymb,
 aps,
]{revtex4-2}
\usepackage{amsmath,amssymb,amsthm}
\usepackage{graphicx}
\usepackage{dcolumn}
\usepackage{bm}
\usepackage{hyperref}

\usepackage{braket}
\usepackage{siunitx}
\newcommand{\dv}{\mathrm{d}} 
\usepackage[caption=false]{subfig}
\begin{document}

\preprint{APS/123-QED}

\title{Cryptographic Security Concerns on Timestamp Sharing via Public Channel in Quantum Key Distribution Systems}

\author{Melis Pahal{\i}}
\email{melis.pahali@ozu.edu.tr}
\affiliation{Electrical and Electronics Engineering Department, \"Ozye\u{g}in University, Turkey}
\author{Utku Tefek}
\email{u.tefek@adsc-create.edu.sg}
\affiliation{Advanced Digital Sciences Center, Singapore}
\author{Kadir Durak}
\email{kadir.durak@ozyegin.edu.tr}
\affiliation{Electrical and Electronics Engineering Department, \"Ozye\u{g}in University, Turkey}%




\date{\today}

\begin{abstract}

Quantum key distribution protocols are known to be vulnerable against a side channel attack that exploits the time difference in detector responses used to obtain key bits. The recommended solution against this timing side channel attack is to use a large time bin width instead of high resolution timing information. Common notion is that using a large bin width reduces the resolution of detector responses, hence supposedly minimizes the information leakage to an eavesdropper. We challenge this conventional wisdom, and demonstrate that increasing the bin width does not monotonically reduce the mutual information between the key bits and the eavesdropper's observation of detector responses. Instead of randomly increasing the bin width, it should be carefully chosen because the mutual information fluctuates with respect to the bin width. We also examined the effect of full width half maximums (FWHMs) of the detectors responses on the mutual information and showed that decreasing the FWHM increases the mutual information. Lastly, the start time of binning is also shown to be important in binning process and the mutual information fluctuates periodically with respect to it.

\end{abstract}

\maketitle


\section{\label{sec:intro}INTRODUCTION}

General scheme of quantum key distribution (QKD) protocols \cite{Ekert1991QuantumCB, Bennett2014QuantumCP} are well known, their security proofs \cite{RenatoRennerSecOfQKD, Zhang2020SecurityPO, SecurityProofTwinFieldQKD, AproofOfSecOfQKD}, side channels of QKD systems \cite{EffOfDetEfficiencyMis, PhaseRemappingAttack, Lydersen:10,10.5555/2840819.2840896, 10.1109/TCAD.2017.2768420, Biswas2021ExperimentalSC, PhysRevA.104.012601}, and the side channel attacks \cite{ExperimentalDemonstrationOfTimeShiftAttack, ExpDOfPhaseRemappingAttack, HackingByTailoredBrightIllumination, FullfieldIOfPerfectEve, ControllingSuperconductingNanowireBright, FakedStatesAttack, Vakhitov2001LargePA, Tang2013SourceAO, Weier2011QuantumEW, Qi2007TimeshiftAI, Yin2020EntanglementbasedSQ} are widely studied topics. In this paper, we study timing side channel which can also be referred as detector efficiency mismatch \cite{EffOfDetEfficiencyMis, Qi2007TimeshiftAI, ExperimentalDemonstrationOfTimeShiftAttack}. 

Before starting a QKD protocol, detection modules need to be characterized. The characterization involves time synchronization between authorized communicating parties. Time synchronization involves determining the path differences between different receivers to the transmitter, and then compensating of these path differences physically or digitally (as a post-process). Path differences in a QKD experimental setup are represented in terms of time differences between detector responses. During QKD, two communicating parties share the basis set in which a photon detection is realized, and the timestamp of this event via a classical channel to determine the coincidences, to generate raw key bits and to compute the quantum bit error rate and/or the amount of violation of Bell's inequality depending on the protocol followed. The timestamps of the events are important in order to determine the coincidences correctly. When the time synchronization is not perfect but realized up to a precision, the coincidences can still be determined correctly, however, a timing side channel attack can occur within that precision. An eavesdropper can estimate the raw key bits, which are required to be kept secret, by simply observing the timestamps and the relative delay between coincidence events in the same basis set. In other words, non-overlapping coincidences in time domain make the values of raw key bits predictable for the eavesdropper. In summary, timing side channel can be understood as the exploitable correlation between the timestamps and the measurement results obtained from quantum channel. Measurement results carry the bit content and therefore the security is relied on them.

In the literature, the recognized solution to this problem is to minimize eavesdropper's information by increasing time bin width for the publicly shared detection times \cite{Lamas-Linares:07}. And only two values of time bin width are examined. However, in this study we examine a range of bin width values and show that arbitrarily increasing the bin width does not necessarily mitigate timing side channel attacks. Instead, the bin width should be carefully chosen depending on the delay between the coincidences.

In this paper, we analyse the timing side channel of a QKD system with imperfect time synchronization such that the delays in coincidences are caused by the imperfect synchronization but not caused by clock drifts.

The paper is organized as follows; a detailed analysis of the mutual information is discussed in section \ref{sec:MutualInfo}, two important parameters in the mutual information, start time of binning and the full width half maximum (FWHM) of detector response, are discussed in sections \ref{sec:StartTimeOfBinning} and \ref{sec:FWHM} respectively, and important findings are summarized in section \ref{sec:Conclusion}.

\section{\label{sec:MutualInfo}MUTUAL INFORMATION}

In an entanglement-based QKD system, two communicating parties Alice and Bob measure incoming photons in their detection modules to obtain raw key bits. A detection module involves a number of non commuting basis sets. A property represented by a discrete variable of the incoming photon is measured in one of non commuting basis sets. For example, the polarization of a photon is measured in one of the detector sets dedicated to $(0^\circ,90^\circ)$ and $(-45^\circ,45^\circ)$ polarization measurements and a bit is obtained. To note that, $(0^\circ,90^\circ)$ constitutes a basis set and each of the two orientations, $0^\circ$ and $90^\circ$, carries a bit content. If entangled photon pairs traveling to Alice's and Bob's detection modules are measured in the same basis set, the measurement outcomes are used as raw key bits. For each pair of bits, it is considered that whether they are raw key bits, violation of Bell's inequality bits or discarded bits according to coincided basis sets among communicating parties. 
Similarly, in a prepare and measure (PaM) QKD system the scenario is similar but the measurement outcomes obtained from all the basis sets are used as raw key bits as long as Alice and Bob' basis sets are matching. Again, the measurement basis sets or bases and timestamps are publicly shared in these systems, which makes PaM protocols also vulnerable against timing side channel attacks. Because any time difference between coincidences obtained from the same basis sets makes those bit contents distinguishable. For the clarity of content, we will continue with the entanglement-based QKD protocol. However, the calculations are valid and can be repeated for PaM protocols also.

\begin{figure}[h!]
    \centering
    \includegraphics[width=270px]{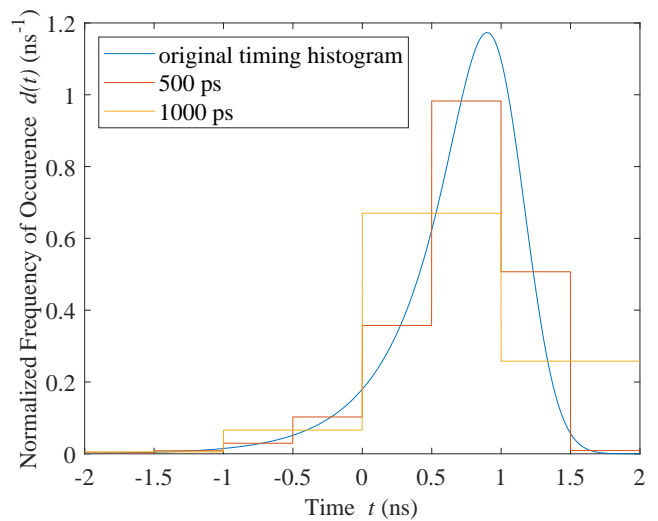}
    \caption{The graph of normalized frequency of occurrence dependence to time. Blue curve is the original timing histogram of the detector, red and yellow curves are the binned versions of the original histogram, and legend represents bin width values.}
    \label{fig:TimingHistogramMakaleden}
\end{figure}

In QKD protocols single photon avalanche diodes (SPADs) are commonly used for detection of photons. For an incident photon, an SPAD outputs a Transistor-Transistor Logic (TTL) signal to create a register having the detection time. There is a time difference between the time that photons falls onto the active area of the SPAD and the generation of a TTL signal. This time difference is a distribution rather than a constant value. This timing histogram is called timing jitter or simply detector response. As the model of a detector response, we work with Eq.~(\ref{eq:PhotonCountingFunc}) \cite{Lamas-Linares:07}, which is an exponentially modified Gaussian distribution
\begin{equation}
    d(t)=\frac{1}{2\tau_e}e^{-\frac{\tau_G}{4\tau_e^2}}\cdot e^{\frac{t-t_0}{\tau_e}} erfc(\frac{t-t_0}{\tau_G}),
    \label{eq:PhotonCountingFunc}
\end{equation}
where `$\cdot$' is convolution product, $\tau_e$ and $\tau_G$ are model parameters and the peak density of $d(t)$ is observed at $t=t_0$. The values for a reference detector are $\tau_e=400\si{\pico\second}$, $\tau_G=290\si{\pico\second}$ and $t_0=1000\si{\pico\second}$. A second detector also has the same parameters except for $t_0$. By changing the value of $t_0$ for the second detector, a $\Delta t_0$ time difference between the two detectors is generated. The profile of the reference detector is the blue curve in Fig.\ref{fig:TimingHistogramMakaleden}. In Fig.\ref{fig:TimingHistogramMakaleden}, the y-axis represents the normalized frequency of occurrence, which is equivalent to the probability density for the original timing histogram.

\begin{figure}[h!]
    \subfloat[]{\includegraphics[width=7.5cm]{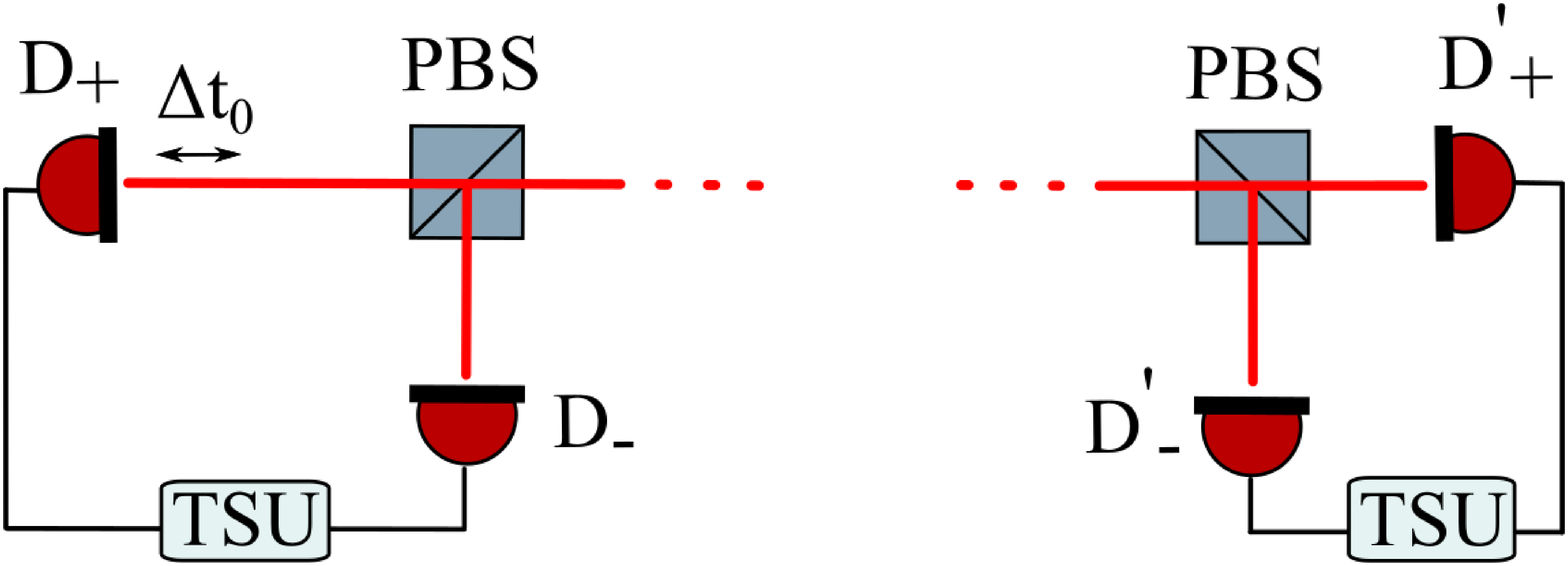}}
    \qquad
    \subfloat[]{\includegraphics[width=6cm]{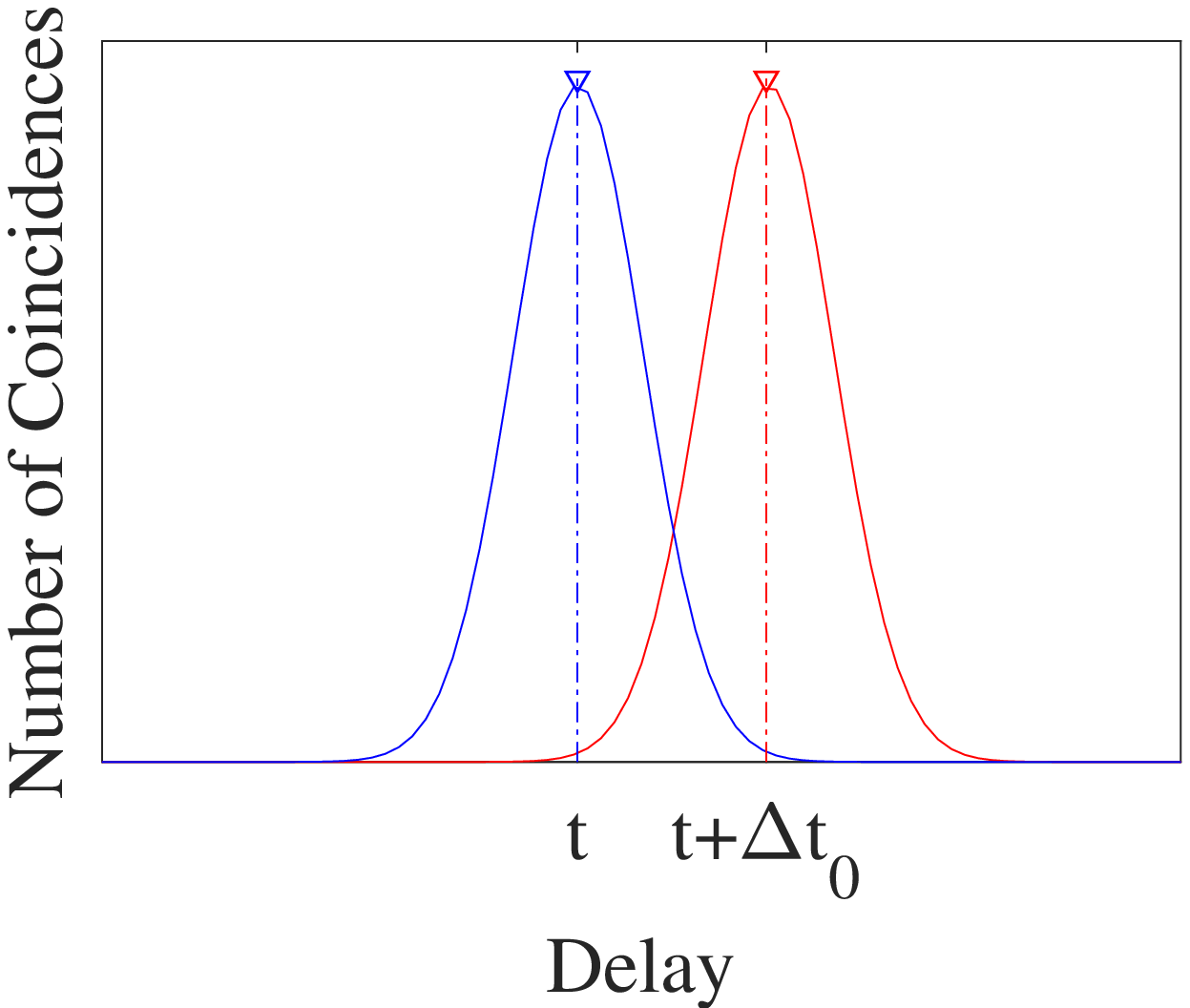} }
    \caption{(a) Schematic representation of the detectors in the same basis sets in Alice's (left) and Bob's (right) sides. PBS is polarizing beam splitter. $+$ and $-$ signs denotes the detectors in transmitted and reflected arms, respectively. TSU is timestamp unit. (b) The cross-correlogram of coincidence events between Alice and Bob when there is an extra optical path in Alice's $D_{+}$ detector. The relative delay between coincidence peaks are due to the extra optical path in transmitted arm on Alice's side. Blue and red peaks represent different delays between $D_{-}$ and $D^{'}_{-}$, and $D_{+}$ and $D^{'}_{+}$, respectively.}
    \label{fig:figure 2}
\end{figure}

Since there are two detectors dedicated to the measurement of raw key bits in one basis set, there are also two different paths a photon can traverse. Each path contains an optical path plus a distance from the start to the peak position of detector response. $\Delta t_0$ is the difference between these two paths. For example, if the raw key bit detection parts of the detection modules at Alice's and Bob's sides are represented schematically as in Fig.\ref{fig:figure 2}(a), the photon that goes to $D_{+}$ travels more than the one goes to $D_{-}$ at Alice's side (left hand-side on Fig.\ref{fig:figure 2}(a)). If the entangled state in the quantum channel between Alice and Bob is $(\ket{00}+\ket{11})/\sqrt{2}$, then the coincidence events coming from the detectors $D_{+}$ and $D^{'}_{+}$, and $D_{-}$ and $D^{'}_{-}$ are seen as separate peaks in a cross-correlogram as shown in the sketch in Fig.\ref{fig:figure 2}(b). The existence of this time difference will cause information leakage to an eavesdropper. This information leakage is quantified with mutual information.

Time bin is the unit time interval used in a QKD system. In every time bin, a measurement is performed in detection modules or not, and one bit of information contributing to a bit string is obtained in each time bin. Time bin width determines the precision of timing histograms and the precision of timestamp information revealed via classical channel by communicating parties. However, there are processes which are binning for discrete time signals and quantization for continuous time signals. In binning process, the value of bin width is redetermined and representative values for normalized frequency of occurrences falling into each time bin are regenerated. In our study, we work with discrete time signals, namely quantized continuous time signals, and we apply the binning process on them. As an example to binning process, in Fig.\ref{fig:TimingHistogramMakaleden} the red and yellow curves are the binned versions of the original timing histogram. The bin width is $500\si{\pico\second}$ in the red curve and $1000\si{\pico\second}$ in the yellow curve.

In the literature, it is known that binning process with large bin width value reduces the mutual information; however, in this study we show that not every increment of the value of bin width gives a reduction in the mutual information. Instead, there is a fluctuation behaviour which is explained in the rest of the paper.

The mutual information $I(X;T)$ between the raw key bit values and detection times can be expressed as in Eq.~(\ref{eq:MutualInfo2}).

\begin{eqnarray}\label{eq:MutualInfo2}
I(X;T)
 &=& \sum_x \int p(x,t) \log \left[\frac{p(x,t)}{p(x)p(t)}\right] \dv t \nonumber\\
 &=& \sum_x \int p(x) p(t \mid x) \log \left[\frac{p(t \mid x)}{p(t)}\right] \dv t \nonumber\\
 &=& \sum_x \int p^0(x) d_x(t) \log \left[\frac{d_x(t)}{\Bar{d}(t)}\right] \dv t \nonumber\\
 &=& \sum_x p^0(x) \int d_x(t) \log d_x(t) \dv t \\& &- \int \Bar{d}(t) \log \Bar{d}(t) \dv t \nonumber
\end{eqnarray}
where the first line follows from definition of mutual information and the third from substituting the relevant distributions. Namely, $p^0(x)$ is the probability mass function of logic bit $x$, $d_x(t)$ is the probability density of taking a click at time $t$ given $x$, and $\Bar{d}(t)$ is the probability density of taking a click at time $t$ from an ensemble of detectors. The last step follows from algebraic manipulations and substituting Eq.~(\ref{eq:MutualInfo3}).
\begin{equation}\label{eq:MutualInfo3}
    \Bar{d}(t)=\sum_x p^0(x)d_x(t)
\end{equation}

\begin{figure}[h!]
    \centering
    \includegraphics[width=270px]{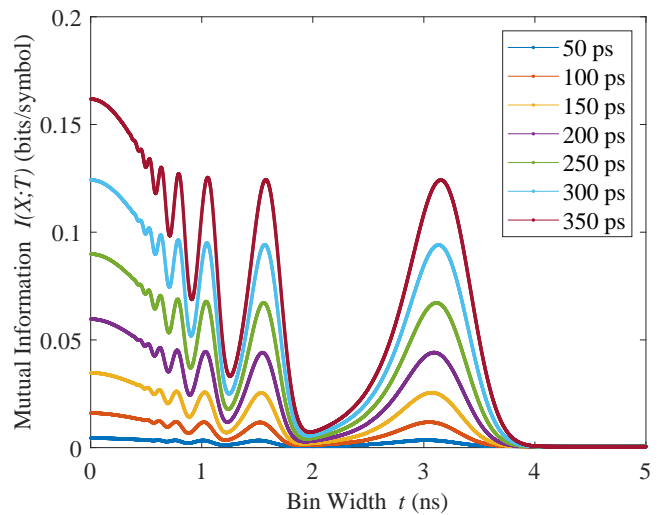}
    \caption{The graph of the mutual information dependence to the bin width and $\Delta t_0$. Legend represents $\Delta t_0$ values.}
    \label{fig:AllBinWidths}
\end{figure}

Mutual information is computed for various $\Delta t_0$ values and it fluctuates as shown in Fig.\ref{fig:AllBinWidths} with varying bin width values. For each bin width value, binning is started at the same point of the original timing histogram so there is no phase difference between the binning processes. 

The fluctuating behaviour of mutual information with increasing bin width indicates that the bin width value should be carefully adjusted and should not arbitrarily increased to reduce the mutual information. This can be realized from the fact that the mutual information for bin width of $1.0\si{\nano\second}$, $1.6\si{\nano\second}$ and $3.2\si{\nano\second}$ are almost the same. 

\section{\label{sec:StartTimeOfBinning}start time of binning}

\begin{figure}[h!]
    \centering
    \includegraphics[width=245px]{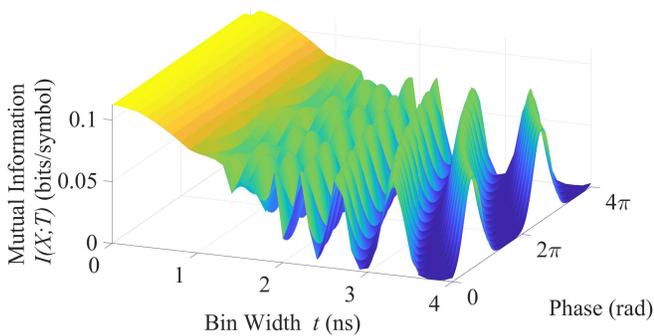}
    \caption{The graph of the mutual information dependence to the start time of binning. The start time of binning is represented in terms of phase.}
    \label{fig:StartTimeOfBinning}
\end{figure}

The index of the time bin at which a photon detection event is registered depends on the start time of binning as well as the bin width. Mutual information changes according to the start time of binning, it is periodic with respect to it for a constant bin width value and the period is equal to the bin width itself. As a result, the start time of binning is also very critical for QKD security. In order to model this behaviour, we use Gaussian distribution for timing histograms of two detectors, in sections \ref{sec:StartTimeOfBinning} and \ref{sec:FWHM}. For illustrating the effect of the start time of binning on the mutual information we considered the following values: $\Delta t_0$ is taken as $350\si{\pico\second}$, the FWHMs of detectors responses are taken as $1\si{\nano\second}$ and the mutual information is calculated for bin width values ranging from $0$ to $4\si{\nano\second}$ in Fig.\ref{fig:StartTimeOfBinning}. In all the mutual information calculations related to different values of bin width, the phase at the start is $0$ and after one bin width the phase becomes $2\pi$. Instead of showing the start time of binning as an axis in Fig.\ref{fig:StartTimeOfBinning}, we chose the phase as a better indicator of periodicity. In order to show the periodicity obviously, the phase is chosen from $0$ to $4\pi$. This graph shows that even carefully chosen bin width is not sufficient for QKD security as the start timing of binning also plays a crucial role.

\section{\label{sec:FWHM}FWHM of detector response}

\begin{figure}[h!]
    \subfloat[]{{\includegraphics[width=125px]{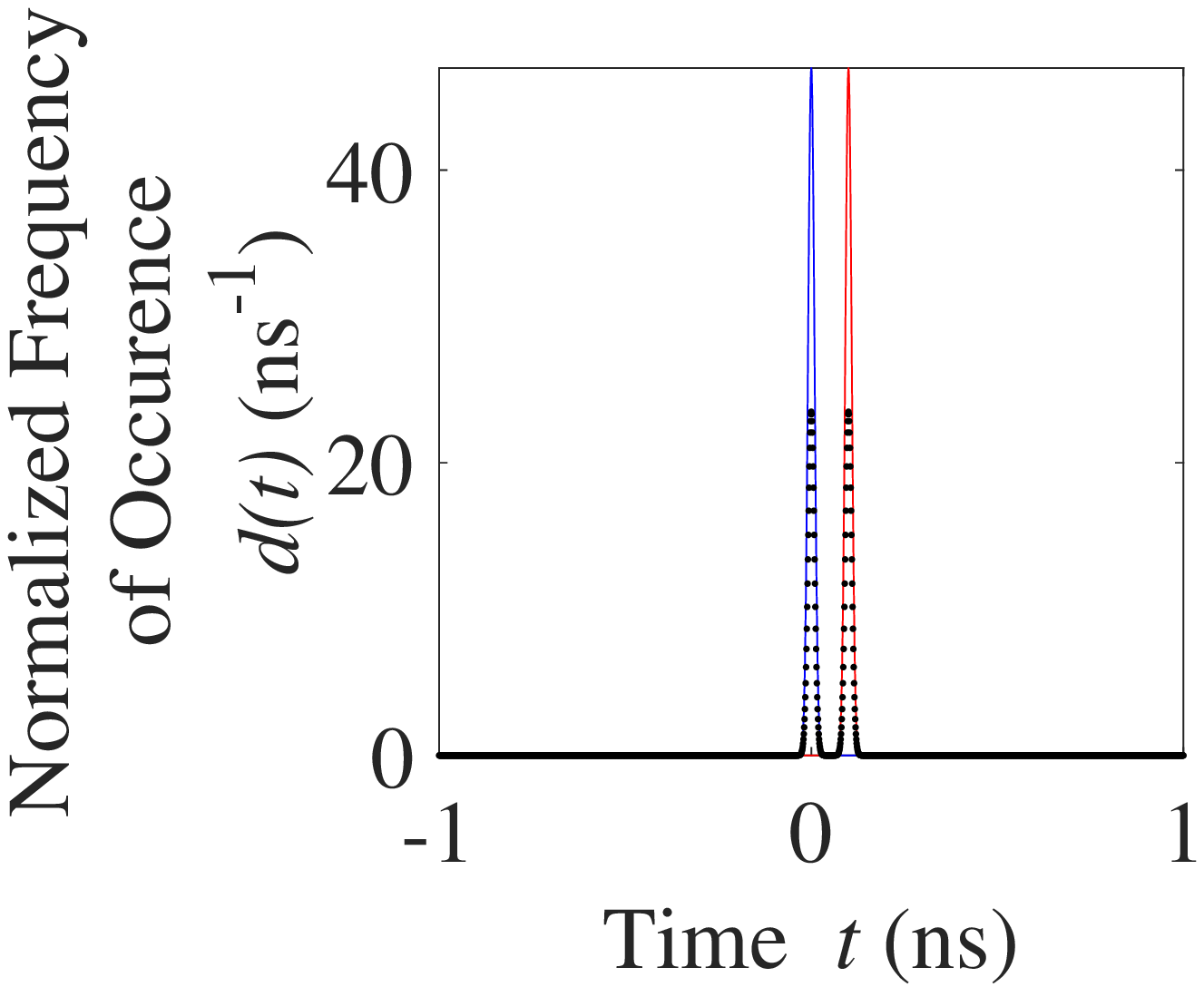} }}
    \subfloat[]{{\includegraphics[width=125px]{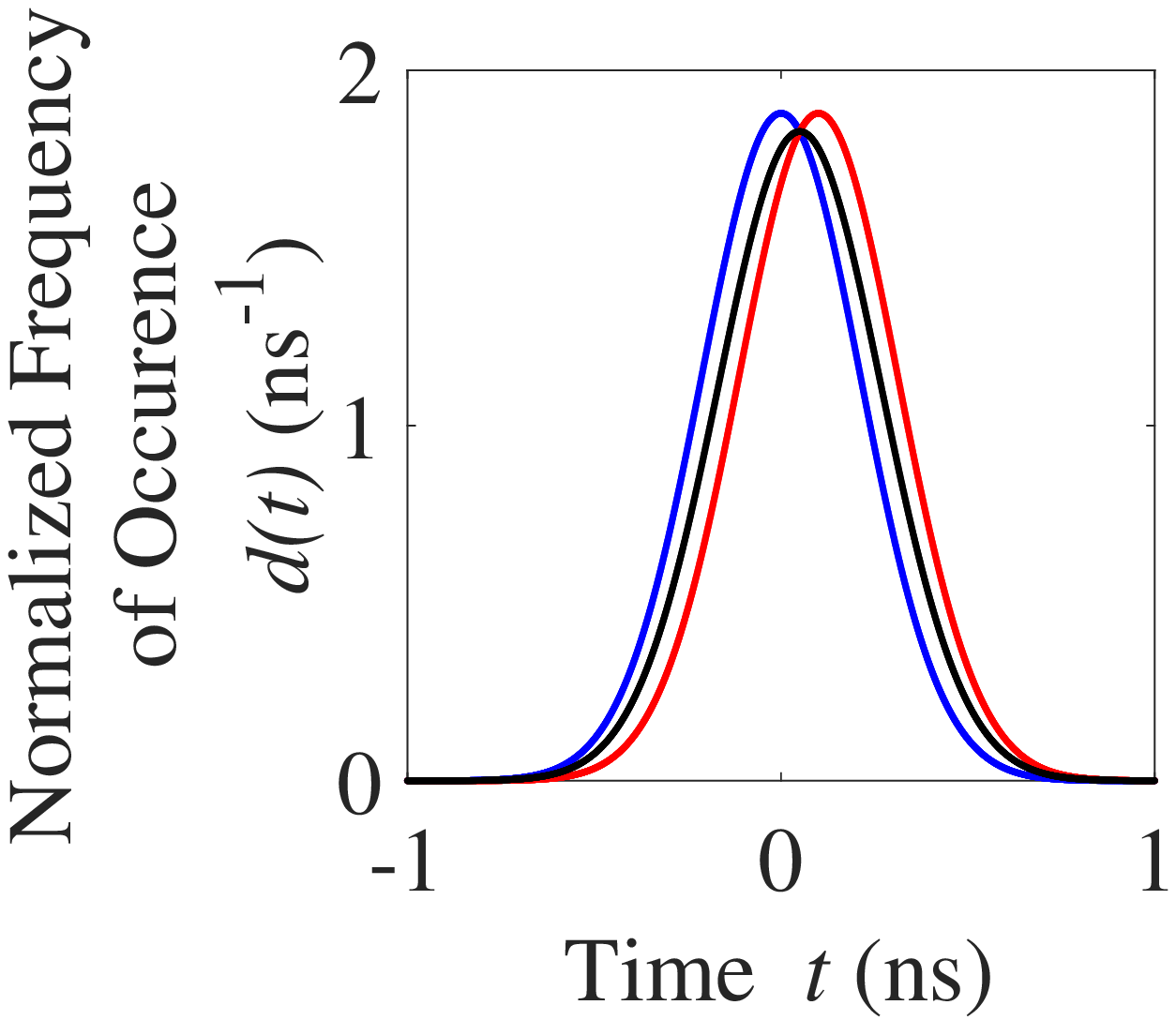} }}
    \caption{The effect of the FWHMs of detectors responses having $\Delta t_0=100\si{\pico\second}$ on the distinguishability of the profiles and hence the values of coincidences. Blue and red curves are two detectors responses. Black curve is the summation of them after multiplied by $0.5$. The distinguishability of the two can be seen in the summation curve. (a) FWHMs $=20\si{\pico\second}$. (b) FWHMs $=500\si{\pico\second}$.}
    \label{fig:fwhmeffect}
\end{figure}

In the quantum industry market, there is a variety of companies offering detectors having low time jitters, which are characterized by FWHMs, as products in order to allow high key rates in QKD implementations. On the other hand, for a constant $\Delta t_0$, when FWHMs of detectors' responses decrease, the overlap portion of their timing histograms decreases and their profiles become distinguishable as can be seen in Fig.\ref{fig:fwhmeffect}, and ultimately the mutual information increases. For this reason, detectors having low time jitters are vulnerable to the timing side channel attack.

\begin{figure}[h!]
    \centering
    \includegraphics[width=245px]{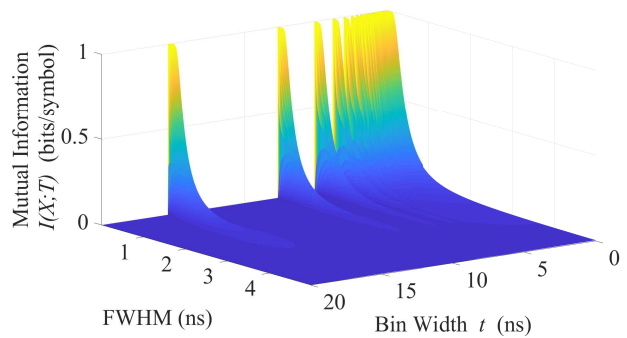}
    \caption{The graph of the mutual information dependence to the FWHMs of detectors responses with varying bin width.}
    \label{fig:FWHM}
\end{figure}

Fig.\ref{fig:FWHM} shows the scan of FWHM of detector response and bin width values for $\Delta t_0=350\si{\pico\second}$. A constant start time of binning is used for varying bin width. It is very critical that when the FWHM of detector response is smaller than $\Delta t_0$, the mutual information is almost $1$. This is a direct consequence of the fact that the coincidence peaks in the cross-correlogram are almost completely distinguishable from the publicly shared timestamps and basis sets when the FWHM of detector response is larger than $\Delta t_0$. The message that should be taken from this graph is that the FWHMs of detectors responses should be negligible compared to $\Delta t_0$. More generally, the coincidence peaks in the cross-correlogram should overlap as mush as possible. This can easily be arranged in a typical QKD setup by a careful physical adjustment of the distances from the PBS to the detectors, and using the detectors the same (or similar) FWHMs of the responses and/or electronic delays. However, an exhaustive search of ways to externally give a delay to one of the coincidence curves is required to prevent a loophole in the security of the QKD system.

The timing side channel mentioned in this work is due to the timestamp and basis set information sharing of Alice and Bob in the public channel. An alternative solution to this problem is that only Alice (Bob) shares the timestamp and basis set information in the public channel and Bob (Alice) checks for the distinguishability in the cross-correlogram peaks of $+$ and $-$ coincidences. After the analysis, Bob eliminates the offset delay in one of the detectors to make the cross-correlogram peaks completely overlapping. After this compensation Bob can share his timestamp and basis set information with Alice in the public channel. In this way, the eavesdropper never learns about the relative delay in coincidences from different detector sets in a basis set.

\section{\label{sec:Conclusion}CONCLUSION}

As a conclusion, in order to reduce or minimize the information leakage to an eavesdropper, choosing the bin width and start timing of the binning values are very important in QKD protocols. Because the mutual information fluctuates with respect to the bin width and the mutual information periodically fluctuates with respect to the start time of binning. Therefore it can be concluded that arbitrarily increasing the bin width is not a precaution to a timing side channel attack. For the security of a QKD system, the bin width, the start time of binning and the ratio of the FWHMs of detectors responses to the time difference between the timing histograms or the coincidence peaks in the cross-correlogram should be carefully adjusted with the characterization of physical system parameters. It is also possible to avoid the timing side channels by allowing one side to compensate for the time delays between cross-correlogram peaks' timing values. 

\begin{acknowledgments}

This research is supported by TÜBİTAK ARDEB, project no 118E991.

\end{acknowledgments}




\bibliography{ms}

\providecommand{\noopsort}[1]{}\providecommand{\singleletter}[1]{#1}%
\begin{thebibliography}{25}%
\makeatletter
\providecommand \@ifxundefined [1]{%
 \@ifx{#1\undefined}
}%
\providecommand \@ifnum [1]{%
 \ifnum #1\expandafter \@firstoftwo
 \else \expandafter \@secondoftwo
 \fi
}%
\providecommand \@ifx [1]{%
 \ifx #1\expandafter \@firstoftwo
 \else \expandafter \@secondoftwo
 \fi
}%
\providecommand \natexlab [1]{#1}%
\providecommand \enquote  [1]{``#1''}%
\providecommand \bibnamefont  [1]{#1}%
\providecommand \bibfnamefont [1]{#1}%
\providecommand \citenamefont [1]{#1}%
\providecommand \href@noop [0]{\@secondoftwo}%
\providecommand \href [0]{\begingroup \@sanitize@url \@href}%
\providecommand \@href[1]{\@@startlink{#1}\@@href}%
\providecommand \@@href[1]{\endgroup#1\@@endlink}%
\providecommand \@sanitize@url [0]{\catcode `\\12\catcode `\$12\catcode
  `\&12\catcode `\#12\catcode `\^12\catcode `\_12\catcode `\%12\relax}%
\providecommand \@@startlink[1]{}%
\providecommand \@@endlink[0]{}%
\providecommand \url  [0]{\begingroup\@sanitize@url \@url }%
\providecommand \@url [1]{\endgroup\@href {#1}{\urlprefix }}%
\providecommand \urlprefix  [0]{URL }%
\providecommand \Eprint [0]{\href }%
\providecommand \doibase [0]{https://doi.org/}%
\providecommand \selectlanguage [0]{\@gobble}%
\providecommand \bibinfo  [0]{\@secondoftwo}%
\providecommand \bibfield  [0]{\@secondoftwo}%
\providecommand \translation [1]{[#1]}%
\providecommand \BibitemOpen [0]{}%
\providecommand \bibitemStop [0]{}%
\providecommand \bibitemNoStop [0]{.\EOS\space}%
\providecommand \EOS [0]{\spacefactor3000\relax}%
\providecommand \BibitemShut  [1]{\csname bibitem#1\endcsname}%
\let\auto@bib@innerbib\@empty
\bibitem [{\citenamefont {Ekert}(1991)}]{Ekert1991QuantumCB}%
  \BibitemOpen
  \bibfield  {author} {\bibinfo {author} {\bibnamefont {Ekert}},\ }\bibfield
  {title} {\bibinfo {title} {Quantum cryptography based on bell's theorem.},\
  }\href@noop {} {\bibfield  {journal} {\bibinfo  {journal} {Physical review
  letters}\ }\textbf {\bibinfo {volume} {67 6}},\ \bibinfo {pages} {661}
  (\bibinfo {year} {1991})}\BibitemShut {NoStop}%
\bibitem [{\citenamefont {Bennett}\ and\ \citenamefont
  {Brassard}(2014)}]{Bennett2014QuantumCP}%
  \BibitemOpen
  \bibfield  {author} {\bibinfo {author} {\bibfnamefont {C.~H.}\ \bibnamefont
  {Bennett}}\ and\ \bibinfo {author} {\bibfnamefont {G.}~\bibnamefont
  {Brassard}},\ }\bibfield  {title} {\bibinfo {title} {Quantum cryptography:
  Public key distribution and coin tossing},\ }\href@noop {} {\bibfield
  {journal} {\bibinfo  {journal} {Theor. Comput. Sci.}\ }\textbf {\bibinfo
  {volume} {560}},\ \bibinfo {pages} {7} (\bibinfo {year} {2014})}\BibitemShut
  {NoStop}%
\bibitem [{\citenamefont {RENNER}(2008)}]{RenatoRennerSecOfQKD}%
  \BibitemOpen
  \bibfield  {author} {\bibinfo {author} {\bibfnamefont {R.}~\bibnamefont
  {RENNER}},\ }\bibfield  {title} {\bibinfo {title} {Security of quantum key
  distribution},\ }\href {https://doi.org/10.1142/S0219749908003256} {\bibfield
   {journal} {\bibinfo  {journal} {International Journal of Quantum
  Information}\ }\textbf {\bibinfo {volume} {06}},\ \bibinfo {pages} {1}
  (\bibinfo {year} {2008})},\ \Eprint
  {https://arxiv.org/abs/https://doi.org/10.1142/S0219749908003256}
  {https://doi.org/10.1142/S0219749908003256} \BibitemShut {NoStop}%
\bibitem [{\citenamefont {Zhang}\ \emph {et~al.}(2020)\citenamefont {Zhang},
  \citenamefont {Coles}, \citenamefont {Winick}, \citenamefont {Lin},\ and\
  \citenamefont {Lutkenhaus}}]{Zhang2020SecurityPO}%
  \BibitemOpen
  \bibfield  {author} {\bibinfo {author} {\bibfnamefont {Y.}~\bibnamefont
  {Zhang}}, \bibinfo {author} {\bibfnamefont {P.~J.}\ \bibnamefont {Coles}},
  \bibinfo {author} {\bibfnamefont {A.~B.}\ \bibnamefont {Winick}}, \bibinfo
  {author} {\bibfnamefont {J.}~\bibnamefont {Lin}},\ and\ \bibinfo {author}
  {\bibfnamefont {N.}~\bibnamefont {Lutkenhaus}},\ }\bibfield  {title}
  {\bibinfo {title} {Security proof of practical quantum key distribution with
  detection-efficiency mismatch},\ }\href@noop {} {\bibfield  {journal}
  {\bibinfo  {journal} {arXiv: Quantum Physics}\ } (\bibinfo {year}
  {2020})}\BibitemShut {NoStop}%
\bibitem [{\citenamefont {Curty}\ \emph {et~al.}(2019)\citenamefont {Curty},
  \citenamefont {Azuma},\ and\ \citenamefont {Lo}}]{SecurityProofTwinFieldQKD}%
  \BibitemOpen
  \bibfield  {author} {\bibinfo {author} {\bibfnamefont {M.}~\bibnamefont
  {Curty}}, \bibinfo {author} {\bibfnamefont {K.}~\bibnamefont {Azuma}},\ and\
  \bibinfo {author} {\bibfnamefont {H.-K.}\ \bibnamefont {Lo}},\ }\bibfield
  {title} {\bibinfo {title} {Simple security proof of twin-field type quantum
  key distribution protocol},\ }\href
  {https://doi.org/10.1038/s41534-019-0175-6} {\bibfield  {journal} {\bibinfo
  {journal} {npj Quantum Information}\ }\textbf {\bibinfo {volume} {5}}
  (\bibinfo {year} {2019})}\BibitemShut {NoStop}%
\bibitem [{\citenamefont {Biham}\ \emph {et~al.}(2006)\citenamefont {Biham},
  \citenamefont {Boyer}, \citenamefont {Boykin}, \citenamefont {Mor},\ and\
  \citenamefont {Roychowdhury}}]{AproofOfSecOfQKD}%
  \BibitemOpen
  \bibfield  {author} {\bibinfo {author} {\bibfnamefont {E.}~\bibnamefont
  {Biham}}, \bibinfo {author} {\bibfnamefont {M.}~\bibnamefont {Boyer}},
  \bibinfo {author} {\bibfnamefont {P.}~\bibnamefont {Boykin}}, \bibinfo
  {author} {\bibfnamefont {T.}~\bibnamefont {Mor}},\ and\ \bibinfo {author}
  {\bibfnamefont {V.}~\bibnamefont {Roychowdhury}},\ }\bibfield  {title}
  {\bibinfo {title} {A proof of the security of quantum key distribution},\
  }\href {https://doi.org/10.1007/s00145-005-0011-3} {\bibfield  {journal}
  {\bibinfo  {journal} {Journal of Cryptology}\ }\textbf {\bibinfo {volume}
  {19}},\ \bibinfo {pages} {381} (\bibinfo {year} {2006})}\BibitemShut
  {NoStop}%
\bibitem [{\citenamefont {Makarov}\ \emph {et~al.}(2005)\citenamefont
  {Makarov}, \citenamefont {Anisimov},\ and\ \citenamefont
  {Skaar}}]{EffOfDetEfficiencyMis}%
  \BibitemOpen
  \bibfield  {author} {\bibinfo {author} {\bibfnamefont {V.}~\bibnamefont
  {Makarov}}, \bibinfo {author} {\bibfnamefont {A.}~\bibnamefont {Anisimov}},\
  and\ \bibinfo {author} {\bibfnamefont {J.}~\bibnamefont {Skaar}},\ }\bibfield
   {title} {\bibinfo {title} {Effects of detector efficiency mismatch on
  security of quantum cryptosystems},\ }\href
  {https://doi.org/10.1103/PHYSREVA.74.022313} {\bibfield  {journal} {\bibinfo
  {journal} {Physical Review A}\ }\textbf {\bibinfo {volume} {74}} (\bibinfo
  {year} {2005})}\BibitemShut {NoStop}%
\bibitem [{\citenamefont {Fung}\ \emph {et~al.}(2007)\citenamefont {Fung},
  \citenamefont {Qi}, \citenamefont {Tamaki},\ and\ \citenamefont
  {Lo}}]{PhaseRemappingAttack}%
  \BibitemOpen
  \bibfield  {author} {\bibinfo {author} {\bibfnamefont {C.-H.~F.}\
  \bibnamefont {Fung}}, \bibinfo {author} {\bibfnamefont {B.}~\bibnamefont
  {Qi}}, \bibinfo {author} {\bibfnamefont {K.}~\bibnamefont {Tamaki}},\ and\
  \bibinfo {author} {\bibfnamefont {H.-K.}\ \bibnamefont {Lo}},\ }\bibfield
  {title} {\bibinfo {title} {Phase-remapping attack in practical
  quantum-key-distribution systems},\ }\href
  {https://doi.org/10.1103/PhysRevA.75.032314} {\bibfield  {journal} {\bibinfo
  {journal} {Phys. Rev. A}\ }\textbf {\bibinfo {volume} {75}},\ \bibinfo
  {pages} {032314} (\bibinfo {year} {2007})}\BibitemShut {NoStop}%
\bibitem [{\citenamefont {Lydersen}\ \emph
  {et~al.}(2010{\natexlab{a}})\citenamefont {Lydersen}, \citenamefont
  {Wiechers}, \citenamefont {Wittmann}, \citenamefont {Elser}, \citenamefont
  {Skaar},\ and\ \citenamefont {Makarov}}]{Lydersen:10}%
  \BibitemOpen
  \bibfield  {author} {\bibinfo {author} {\bibfnamefont {L.}~\bibnamefont
  {Lydersen}}, \bibinfo {author} {\bibfnamefont {C.}~\bibnamefont {Wiechers}},
  \bibinfo {author} {\bibfnamefont {C.}~\bibnamefont {Wittmann}}, \bibinfo
  {author} {\bibfnamefont {D.}~\bibnamefont {Elser}}, \bibinfo {author}
  {\bibfnamefont {J.}~\bibnamefont {Skaar}},\ and\ \bibinfo {author}
  {\bibfnamefont {V.}~\bibnamefont {Makarov}},\ }\bibfield  {title} {\bibinfo
  {title} {Thermal blinding of gated detectors in quantum cryptography},\
  }\href {https://doi.org/10.1364/OE.18.027938} {\bibfield  {journal} {\bibinfo
   {journal} {Opt. Express}\ }\textbf {\bibinfo {volume} {18}},\ \bibinfo
  {pages} {27938} (\bibinfo {year} {2010}{\natexlab{a}})}\BibitemShut {NoStop}%
\bibitem [{\citenamefont {Mao}\ \emph {et~al.}(2015)\citenamefont {Mao},
  \citenamefont {Hu}, \citenamefont {Althoff}, \citenamefont {Matai},
  \citenamefont {Oberg}, \citenamefont {Mu}, \citenamefont {Sherwood},\ and\
  \citenamefont {Kastner}}]{10.5555/2840819.2840896}%
  \BibitemOpen
  \bibfield  {author} {\bibinfo {author} {\bibfnamefont {B.}~\bibnamefont
  {Mao}}, \bibinfo {author} {\bibfnamefont {W.}~\bibnamefont {Hu}}, \bibinfo
  {author} {\bibfnamefont {A.}~\bibnamefont {Althoff}}, \bibinfo {author}
  {\bibfnamefont {J.}~\bibnamefont {Matai}}, \bibinfo {author} {\bibfnamefont
  {J.}~\bibnamefont {Oberg}}, \bibinfo {author} {\bibfnamefont
  {D.}~\bibnamefont {Mu}}, \bibinfo {author} {\bibfnamefont {T.}~\bibnamefont
  {Sherwood}},\ and\ \bibinfo {author} {\bibfnamefont {R.}~\bibnamefont
  {Kastner}},\ }\bibfield  {title} {\bibinfo {title} {Quantifying timing-based
  information flow in cryptographic hardware},\ }in\ \href@noop {} {\emph
  {\bibinfo {booktitle} {Proceedings of the IEEE/ACM International Conference
  on Computer-Aided Design}}},\ \bibinfo {series and number} {ICCAD '15}\
  (\bibinfo  {publisher} {IEEE Press},\ \bibinfo {year} {2015})\ p.\ \bibinfo
  {pages} {552–559}\BibitemShut {NoStop}%
\bibitem [{\citenamefont {Mao}\ \emph {et~al.}(2018)\citenamefont {Mao},
  \citenamefont {Hu}, \citenamefont {Althoff}, \citenamefont {Matai},
  \citenamefont {Tai}, \citenamefont {Mu}, \citenamefont {Sherwood},\ and\
  \citenamefont {Kastner}}]{10.1109/TCAD.2017.2768420}%
  \BibitemOpen
  \bibfield  {author} {\bibinfo {author} {\bibfnamefont {B.}~\bibnamefont
  {Mao}}, \bibinfo {author} {\bibfnamefont {W.}~\bibnamefont {Hu}}, \bibinfo
  {author} {\bibfnamefont {A.}~\bibnamefont {Althoff}}, \bibinfo {author}
  {\bibfnamefont {J.}~\bibnamefont {Matai}}, \bibinfo {author} {\bibfnamefont
  {Y.}~\bibnamefont {Tai}}, \bibinfo {author} {\bibfnamefont {D.}~\bibnamefont
  {Mu}}, \bibinfo {author} {\bibfnamefont {T.}~\bibnamefont {Sherwood}},\ and\
  \bibinfo {author} {\bibfnamefont {R.}~\bibnamefont {Kastner}},\ }\bibfield
  {title} {\bibinfo {title} {Quantitative analysis of timing channel security
  in cryptographic hardware design},\ }\href
  {https://doi.org/10.1109/TCAD.2017.2768420} {\bibfield  {journal} {\bibinfo
  {journal} {Trans. Comp.-Aided Des. Integ. Cir. Sys.}\ }\textbf {\bibinfo
  {volume} {37}},\ \bibinfo {pages} {1719–1732} (\bibinfo {year}
  {2018})}\BibitemShut {NoStop}%
\bibitem [{\citenamefont {Biswas}\ \emph {et~al.}(2021)\citenamefont {Biswas},
  \citenamefont {Banerji}, \citenamefont {Chandravanshi}, \citenamefont
  {Kumar},\ and\ \citenamefont {Singh}}]{Biswas2021ExperimentalSC}%
  \BibitemOpen
  \bibfield  {author} {\bibinfo {author} {\bibfnamefont {A.~K.}\ \bibnamefont
  {Biswas}}, \bibinfo {author} {\bibfnamefont {A.}~\bibnamefont {Banerji}},
  \bibinfo {author} {\bibfnamefont {P.}~\bibnamefont {Chandravanshi}}, \bibinfo
  {author} {\bibfnamefont {R.}~\bibnamefont {Kumar}},\ and\ \bibinfo {author}
  {\bibfnamefont {R.~P.}\ \bibnamefont {Singh}},\ }\bibfield  {title} {\bibinfo
  {title} {Experimental side channel analysis of bb84 qkd source},\ }\href@noop
  {} {\bibfield  {journal} {\bibinfo  {journal} {IEEE Journal of Quantum
  Electronics}\ }\textbf {\bibinfo {volume} {57}},\ \bibinfo {pages} {1}
  (\bibinfo {year} {2021})}\BibitemShut {NoStop}%
\bibitem [{\citenamefont {Duplinskiy}\ and\ \citenamefont
  {Sych}(2021)}]{PhysRevA.104.012601}%
  \BibitemOpen
  \bibfield  {author} {\bibinfo {author} {\bibfnamefont {A.}~\bibnamefont
  {Duplinskiy}}\ and\ \bibinfo {author} {\bibfnamefont {D.}~\bibnamefont
  {Sych}},\ }\bibfield  {title} {\bibinfo {title} {Bounding passive
  light-source side channels in quantum key distribution via hong-ou-mandel
  interference},\ }\href {https://doi.org/10.1103/PhysRevA.104.012601}
  {\bibfield  {journal} {\bibinfo  {journal} {Phys. Rev. A}\ }\textbf {\bibinfo
  {volume} {104}},\ \bibinfo {pages} {012601} (\bibinfo {year}
  {2021})}\BibitemShut {NoStop}%
\bibitem [{\citenamefont {Zhao}\ \emph {et~al.}(2008)\citenamefont {Zhao},
  \citenamefont {Fung}, \citenamefont {Qi}, \citenamefont {Chen},\ and\
  \citenamefont {Lo}}]{ExperimentalDemonstrationOfTimeShiftAttack}%
  \BibitemOpen
  \bibfield  {author} {\bibinfo {author} {\bibfnamefont {Y.}~\bibnamefont
  {Zhao}}, \bibinfo {author} {\bibfnamefont {C.-H.~F.}\ \bibnamefont {Fung}},
  \bibinfo {author} {\bibfnamefont {B.}~\bibnamefont {Qi}}, \bibinfo {author}
  {\bibfnamefont {C.}~\bibnamefont {Chen}},\ and\ \bibinfo {author}
  {\bibfnamefont {H.-K.}\ \bibnamefont {Lo}},\ }\bibfield  {title} {\bibinfo
  {title} {Quantum hacking: Experimental demonstration of time-shift attack
  against practical quantum-key-distribution systems},\ }\href
  {https://doi.org/10.1103/PhysRevA.78.042333} {\bibfield  {journal} {\bibinfo
  {journal} {Phys. Rev. A}\ }\textbf {\bibinfo {volume} {78}},\ \bibinfo
  {pages} {042333} (\bibinfo {year} {2008})}\BibitemShut {NoStop}%
\bibitem [{\citenamefont {Xu}\ \emph {et~al.}(2010)\citenamefont {Xu},
  \citenamefont {Qi},\ and\ \citenamefont {Lo}}]{ExpDOfPhaseRemappingAttack}%
  \BibitemOpen
  \bibfield  {author} {\bibinfo {author} {\bibfnamefont {F.}~\bibnamefont
  {Xu}}, \bibinfo {author} {\bibfnamefont {B.}~\bibnamefont {Qi}},\ and\
  \bibinfo {author} {\bibfnamefont {H.}~\bibnamefont {Lo}},\ }\bibfield
  {title} {\bibinfo {title} {Experimental demonstration of phase-remapping
  attack in a practical quantum key distribution system},\ }\href@noop {}
  {\bibfield  {journal} {\bibinfo  {journal} {New Journal of Physics}\ }\textbf
  {\bibinfo {volume} {12}},\ \bibinfo {pages} {113026} (\bibinfo {year}
  {2010})}\BibitemShut {NoStop}%
\bibitem [{\citenamefont {Lydersen}\ \emph
  {et~al.}(2010{\natexlab{b}})\citenamefont {Lydersen}, \citenamefont
  {Wiechers}, \citenamefont {Wittmann}, \citenamefont {Elser}, \citenamefont
  {Skaar},\ and\ \citenamefont
  {Makarov}}]{HackingByTailoredBrightIllumination}%
  \BibitemOpen
  \bibfield  {author} {\bibinfo {author} {\bibfnamefont {L.}~\bibnamefont
  {Lydersen}}, \bibinfo {author} {\bibfnamefont {C.}~\bibnamefont {Wiechers}},
  \bibinfo {author} {\bibfnamefont {C.}~\bibnamefont {Wittmann}}, \bibinfo
  {author} {\bibfnamefont {D.}~\bibnamefont {Elser}}, \bibinfo {author}
  {\bibfnamefont {J.}~\bibnamefont {Skaar}},\ and\ \bibinfo {author}
  {\bibfnamefont {V.}~\bibnamefont {Makarov}},\ }\bibfield  {title} {\bibinfo
  {title} {Hacking commercial quantum cryptography systems by tailored bright
  illumination},\ }\href@noop {} {\bibfield  {journal} {\bibinfo  {journal}
  {Nature Photonics}\ }\textbf {\bibinfo {volume} {4}},\ \bibinfo {pages} {686}
  (\bibinfo {year} {2010}{\natexlab{b}})}\BibitemShut {NoStop}%
\bibitem [{\citenamefont {Gerhardt}\ \emph {et~al.}(2011)\citenamefont
  {Gerhardt}, \citenamefont {Liu}, \citenamefont {Lamas-Linares}, \citenamefont
  {Skaar}, \citenamefont {Kurtsiefer},\ and\ \citenamefont
  {Makarov}}]{FullfieldIOfPerfectEve}%
  \BibitemOpen
  \bibfield  {author} {\bibinfo {author} {\bibfnamefont {I.}~\bibnamefont
  {Gerhardt}}, \bibinfo {author} {\bibfnamefont {Q.}~\bibnamefont {Liu}},
  \bibinfo {author} {\bibfnamefont {A.}~\bibnamefont {Lamas-Linares}}, \bibinfo
  {author} {\bibfnamefont {J.}~\bibnamefont {Skaar}}, \bibinfo {author}
  {\bibfnamefont {C.}~\bibnamefont {Kurtsiefer}},\ and\ \bibinfo {author}
  {\bibfnamefont {V.}~\bibnamefont {Makarov}},\ }\bibfield  {title} {\bibinfo
  {title} {Full-field implementation of a perfect eavesdropper on a quantum
  cryptography system.},\ }\href@noop {} {\bibfield  {journal} {\bibinfo
  {journal} {Nature communications}\ }\textbf {\bibinfo {volume} {2}},\
  \bibinfo {pages} {349} (\bibinfo {year} {2011})}\BibitemShut {NoStop}%
\bibitem [{\citenamefont {Lydersen}\ \emph {et~al.}(2011)\citenamefont
  {Lydersen}, \citenamefont {Akhlaghi}, \citenamefont {Majedi}, \citenamefont
  {Skaar},\ and\ \citenamefont
  {Makarov}}]{ControllingSuperconductingNanowireBright}%
  \BibitemOpen
  \bibfield  {author} {\bibinfo {author} {\bibfnamefont {L.}~\bibnamefont
  {Lydersen}}, \bibinfo {author} {\bibfnamefont {M.~K.}\ \bibnamefont
  {Akhlaghi}}, \bibinfo {author} {\bibfnamefont {A.~H.}\ \bibnamefont
  {Majedi}}, \bibinfo {author} {\bibfnamefont {J.}~\bibnamefont {Skaar}},\ and\
  \bibinfo {author} {\bibfnamefont {V.}~\bibnamefont {Makarov}},\ }\bibfield
  {title} {\bibinfo {title} {Controlling a superconducting nanowire
  single-photon detector using tailored bright illumination},\ }\href@noop {}
  {\bibfield  {journal} {\bibinfo  {journal} {New Journal of Physics}\ }\textbf
  {\bibinfo {volume} {13}},\ \bibinfo {pages} {113042} (\bibinfo {year}
  {2011})}\BibitemShut {NoStop}%
\bibitem [{\citenamefont {Makarov}\ and\ \citenamefont
  {Hjelme}(2005)}]{FakedStatesAttack}%
  \BibitemOpen
  \bibfield  {author} {\bibinfo {author} {\bibfnamefont {V.}~\bibnamefont
  {Makarov}}\ and\ \bibinfo {author} {\bibfnamefont {D.}~\bibnamefont
  {Hjelme}},\ }\bibfield  {title} {\bibinfo {title} {Faked states attack on
  quantum cryptosystems},\ }\href@noop {} {\bibfield  {journal} {\bibinfo
  {journal} {Journal of Modern Optics}\ }\textbf {\bibinfo {volume} {52}},\
  \bibinfo {pages} {691 } (\bibinfo {year} {2005})}\BibitemShut {NoStop}%
\bibitem [{\citenamefont {Vakhitov}\ \emph {et~al.}(2001)\citenamefont
  {Vakhitov}, \citenamefont {Makarov},\ and\ \citenamefont
  {Hjelme}}]{Vakhitov2001LargePA}%
  \BibitemOpen
  \bibfield  {author} {\bibinfo {author} {\bibfnamefont {A.}~\bibnamefont
  {Vakhitov}}, \bibinfo {author} {\bibfnamefont {V.}~\bibnamefont {Makarov}},\
  and\ \bibinfo {author} {\bibfnamefont {D.}~\bibnamefont {Hjelme}},\
  }\bibfield  {title} {\bibinfo {title} {Large pulse attack as a method of
  conventional optical eavesdropping in quantum cryptography},\ }\href@noop {}
  {\bibfield  {journal} {\bibinfo  {journal} {Journal of Modern Optics}\
  }\textbf {\bibinfo {volume} {48}},\ \bibinfo {pages} {2023 } (\bibinfo {year}
  {2001})}\BibitemShut {NoStop}%
\bibitem [{\citenamefont {Tang}\ \emph {et~al.}(2013)\citenamefont {Tang},
  \citenamefont {Yin}, \citenamefont {Ma}, \citenamefont {Fung}, \citenamefont
  {Liu}, \citenamefont {Yong}, \citenamefont {Chen}, \citenamefont {Peng},
  \citenamefont {Chen},\ and\ \citenamefont {Pan}}]{Tang2013SourceAO}%
  \BibitemOpen
  \bibfield  {author} {\bibinfo {author} {\bibfnamefont {Y.-L.}\ \bibnamefont
  {Tang}}, \bibinfo {author} {\bibfnamefont {H.-L.}\ \bibnamefont {Yin}},
  \bibinfo {author} {\bibfnamefont {X.}~\bibnamefont {Ma}}, \bibinfo {author}
  {\bibfnamefont {C.}~\bibnamefont {Fung}}, \bibinfo {author} {\bibfnamefont
  {Y.}~\bibnamefont {Liu}}, \bibinfo {author} {\bibfnamefont {H.-L.}\
  \bibnamefont {Yong}}, \bibinfo {author} {\bibfnamefont {T.-Y.}\ \bibnamefont
  {Chen}}, \bibinfo {author} {\bibfnamefont {C.-Z.}\ \bibnamefont {Peng}},
  \bibinfo {author} {\bibfnamefont {Z.-B.}\ \bibnamefont {Chen}},\ and\
  \bibinfo {author} {\bibfnamefont {J.}~\bibnamefont {Pan}},\ }\bibfield
  {title} {\bibinfo {title} {Source attack of decoy-state quantum key
  distribution using phase information},\ }\href@noop {} {\bibfield  {journal}
  {\bibinfo  {journal} {Physical Review A}\ }\textbf {\bibinfo {volume} {88}}
  (\bibinfo {year} {2013})}\BibitemShut {NoStop}%
\bibitem [{\citenamefont {Weier}\ \emph {et~al.}(2011)\citenamefont {Weier},
  \citenamefont {Krauss}, \citenamefont {Rau}, \citenamefont {Fuerst},
  \citenamefont {Nauerth},\ and\ \citenamefont
  {Weinfurter}}]{Weier2011QuantumEW}%
  \BibitemOpen
  \bibfield  {author} {\bibinfo {author} {\bibfnamefont {H.}~\bibnamefont
  {Weier}}, \bibinfo {author} {\bibfnamefont {H.}~\bibnamefont {Krauss}},
  \bibinfo {author} {\bibfnamefont {M.}~\bibnamefont {Rau}}, \bibinfo {author}
  {\bibfnamefont {M.}~\bibnamefont {Fuerst}}, \bibinfo {author} {\bibfnamefont
  {S.}~\bibnamefont {Nauerth}},\ and\ \bibinfo {author} {\bibfnamefont
  {H.}~\bibnamefont {Weinfurter}},\ }\bibfield  {title} {\bibinfo {title}
  {Quantum eavesdropping without interception: an attack exploiting the dead
  time of single-photon detectors},\ }\href@noop {} {\bibfield  {journal}
  {\bibinfo  {journal} {New Journal of Physics}\ }\textbf {\bibinfo {volume}
  {13}},\ \bibinfo {pages} {073024} (\bibinfo {year} {2011})}\BibitemShut
  {NoStop}%
\bibitem [{\citenamefont {Qi}\ \emph {et~al.}(2007)\citenamefont {Qi},
  \citenamefont {Fung}, \citenamefont {Lo},\ and\ \citenamefont
  {Ma}}]{Qi2007TimeshiftAI}%
  \BibitemOpen
  \bibfield  {author} {\bibinfo {author} {\bibfnamefont {B.}~\bibnamefont
  {Qi}}, \bibinfo {author} {\bibfnamefont {C.}~\bibnamefont {Fung}}, \bibinfo
  {author} {\bibfnamefont {H.}~\bibnamefont {Lo}},\ and\ \bibinfo {author}
  {\bibfnamefont {X.}~\bibnamefont {Ma}},\ }\bibfield  {title} {\bibinfo
  {title} {Time-shift attack in practical quantum cryptosystems},\ }\href@noop
  {} {\bibfield  {journal} {\bibinfo  {journal} {Quantum Inf. Comput.}\
  }\textbf {\bibinfo {volume} {7}},\ \bibinfo {pages} {73} (\bibinfo {year}
  {2007})}\BibitemShut {NoStop}%
\bibitem [{\citenamefont {Yin}\ \emph {et~al.}(2020)\citenamefont {Yin},
  \citenamefont {Li}, \citenamefont {Liao}, \citenamefont {Yang}, \citenamefont
  {Cao}, \citenamefont {Zhang}, \citenamefont {Ren}, \citenamefont {Cai},
  \citenamefont {Liu}, \citenamefont {Li}, \citenamefont {Shu}, \citenamefont
  {Huang}, \citenamefont {Deng}, \citenamefont {Li}, \citenamefont {Zhang},
  \citenamefont {Liu}, \citenamefont {Chen}, \citenamefont {Lu}, \citenamefont
  {bin Wang}, \citenamefont {Xu}, \citenamefont {Wang}, \citenamefont {Peng},
  \citenamefont {Ekert},\ and\ \citenamefont
  {Pan}}]{Yin2020EntanglementbasedSQ}%
  \BibitemOpen
  \bibfield  {author} {\bibinfo {author} {\bibfnamefont {J.}~\bibnamefont
  {Yin}}, \bibinfo {author} {\bibfnamefont {Y.-H.}\ \bibnamefont {Li}},
  \bibinfo {author} {\bibfnamefont {S.}~\bibnamefont {Liao}}, \bibinfo {author}
  {\bibfnamefont {M.}~\bibnamefont {Yang}}, \bibinfo {author} {\bibfnamefont
  {Y.}~\bibnamefont {Cao}}, \bibinfo {author} {\bibfnamefont {L.}~\bibnamefont
  {Zhang}}, \bibinfo {author} {\bibfnamefont {J.-G.}\ \bibnamefont {Ren}},
  \bibinfo {author} {\bibfnamefont {W.}~\bibnamefont {Cai}}, \bibinfo {author}
  {\bibfnamefont {W.}~\bibnamefont {Liu}}, \bibinfo {author} {\bibfnamefont
  {S.-L.}\ \bibnamefont {Li}}, \bibinfo {author} {\bibfnamefont
  {R.}~\bibnamefont {Shu}}, \bibinfo {author} {\bibfnamefont {Y.}~\bibnamefont
  {Huang}}, \bibinfo {author} {\bibfnamefont {L.}~\bibnamefont {Deng}},
  \bibinfo {author} {\bibfnamefont {L.}~\bibnamefont {Li}}, \bibinfo {author}
  {\bibfnamefont {Q.}~\bibnamefont {Zhang}}, \bibinfo {author} {\bibfnamefont
  {N.}~\bibnamefont {Liu}}, \bibinfo {author} {\bibfnamefont {Y.}~\bibnamefont
  {Chen}}, \bibinfo {author} {\bibfnamefont {C.-Y.}\ \bibnamefont {Lu}},
  \bibinfo {author} {\bibfnamefont {X.}~\bibnamefont {bin Wang}}, \bibinfo
  {author} {\bibfnamefont {F.}~\bibnamefont {Xu}}, \bibinfo {author}
  {\bibfnamefont {J.-Y.}\ \bibnamefont {Wang}}, \bibinfo {author}
  {\bibfnamefont {C.-Z.}\ \bibnamefont {Peng}}, \bibinfo {author}
  {\bibfnamefont {A.}~\bibnamefont {Ekert}},\ and\ \bibinfo {author}
  {\bibfnamefont {J.}~\bibnamefont {Pan}},\ }\bibfield  {title} {\bibinfo
  {title} {Entanglement-based secure quantum cryptography over 1,120
  kilometres},\ }\href@noop {} {\bibfield  {journal} {\bibinfo  {journal}
  {Nature}\ }\textbf {\bibinfo {volume} {582}},\ \bibinfo {pages} {501}
  (\bibinfo {year} {2020})}\BibitemShut {NoStop}%
\bibitem [{\citenamefont {Lamas-Linares}\ and\ \citenamefont
  {Kurtsiefer}(2007)}]{Lamas-Linares:07}%
  \BibitemOpen
  \bibfield  {author} {\bibinfo {author} {\bibfnamefont {A.}~\bibnamefont
  {Lamas-Linares}}\ and\ \bibinfo {author} {\bibfnamefont {C.}~\bibnamefont
  {Kurtsiefer}},\ }\bibfield  {title} {\bibinfo {title} {Breaking a quantum key
  distribution system through a timing side channel},\ }\href
  {https://doi.org/10.1364/OE.15.009388} {\bibfield  {journal} {\bibinfo
  {journal} {Opt. Express}\ }\textbf {\bibinfo {volume} {15}},\ \bibinfo
  {pages} {9388} (\bibinfo {year} {2007})}\BibitemShut {NoStop}%
\end{thebibliography}%

\end{document}